\documentstyle[tighten,preprint,aps,amsfonts]{revtex}
\DeclareMathSymbol{\shortparallel}  {\mathrel}{AMSb}{"71}
\DeclareMathSymbol{\restriction}  {\mathrel}{AMSa}{"16}
\newcommand{\Tr}{{\rm Tr\,}}
\newcommand{\r}{\varrho}
\newcommand{\dr}{{\,\rm d}\varrho\,}

\newcommand{\dmu}{{\,\rm d}\mu\,}
\newcommand{\du}{{\,\rm d}u\,}
\newcommand{\dP}[1]{{\rm d}p_#1}
\newcommand{\de}[1]{{\rm d}e_#1}
\begin{document}
\draft
\title{Note on Explicit Formulae for the Bures Metric}
\author{J.~Dittmann\thanks{
e-mail correspondence: {\tt dittmann@mathematik.uni-leipzig.de}}}
\address{Mathematisches Institut, Universit\"at Leipzig\\
Augustusplatz 10/11, 04109 Leipzig,
Germany}
\date{August 24, 1998}
\maketitle
\begin{abstract}
The aim of this paper is to derive explicit formulae for
the Riemannian Bures metric $g$
on the manifold ${\cal D}$ of (finite dimensional) nondegenerate density 
matrices $\r$. 
The computation of the Bures metric using these equations
does not require any diagonalization procedure.
The first  equations we give  are, 
essentially,  of the form $g=\sum a_{ij}\Tr \dr\r^{i-1}\dr\r^{j-1}$, where 
$a_{i,j}$ is a  matrix of invariants of $\r$. 
A further   formula,
$g=\sum c_{ij}\,\dP{i}\otimes\dP{j}+
\sum  b_{ij}\Tr \dr\r^{i-1}\dr\r^{j-1}$, 
is adapted to the local orthogonal decomposition 
${\cal D}\approx{\Bbb R}^n\times {}^{{\rm U}(n)}/_{{\rm T}^n}$ at generic 
points. 
\end{abstract}
\pacs{PACS Numbers 03.65.Bz, 02.40.-k, 02.40.Ky}

\section{Introduction}
Let $\cal D$ be the manifold of all nondegenerate, positive, Hermitian
 $n{\times}n$-matrices. The  tangent space  ${\rm T}_\r\cal D$ 
consists of all Hermitian $n{\times}n$-matrices and the Riemannian Bures 
metric on $\cal D$ is defined by, \cite{Uh92b},
\begin{equation}\label{metric}
g_\r(Y',Y)=\frac{1}{2}\Tr Y'X\,,\qquad Y,Y'\in{\rm T}_\r\cal 
D\,,
\end{equation}
where $X$ is the (unique) solution of $\r X+X\r=Y$. The submanifold 
of trace one matrices is the  space of so called completely entangled 
mixed states of a finite dimensional quantum system. 
This metric appears quite naturally on the background of purifications 
of mixed states and is used in quantum information theory to describe the 
statistical distance of mixed states. 
It is an extremal monotone metric and 
seems to be quite distinguished for several mathematical and physical 
reasons, see e.~g.~\cite{UD}. It should be mentioned, that (\ref{metric})
defines also a metric on manifolds of stable range densities, but we will 
deal with the maximal range, only.
 
Several formulae and approaches for computing the Bures metric  were given for low 
dimensions, e.~g.~\cite{Uh92b,Hu92b,Di93a,Twam,Slater}.
If $|\alpha\rangle$, $\alpha=1,\dots,n$, are eigenvectors of $\r$ with 
eigenvalues $\lambda_\alpha$, then a simple 
calculation shows that (\ref{metric}) yields,\cite{Hu92b},
$$g_\r(Y,Y)=\frac{1}{2}\sum_{\alpha,\beta}
\frac{|\langle\alpha|Y|\beta\rangle|^2}{\lambda_\alpha+\lambda_\beta}\,,
$$
but such a formula we will not accept as explicit, because it 
needs the knowledge of eigenvalues. 

The aim of this note is to provide  several equations for 
computing the Bures metric in any finite dimension using matrix 
products, determinants and traces, only. 
Section II provides  expressions for the Bures metric based, 
essentially, on results of the theory of matrix equations. In 
Section III we give an equation adapted to the local isometric 
decomposition of the manifold $\cal D$.

\smallskip
\noindent{\bf Notations:}
The following quantities depend on  a positive, Hermitian 
$n\times n$ - matrix $\r$. In order 
to simplify the notation the dependence on $\r$ will be suppressed. 
By $\lambda_1\leq\dots\leq\lambda_n$ we denote the eigenvalues of $\r$ and 
by $\Lambda$ the diagonal matrix of the $\lambda_i$-s. Moreover,
$V$ will be the Vandermonde matrix 
$\left(\lambda_i^{j-1}\right)$.
Operators acting on matrices are denoted by bold letters, especially,
if not indicated otherwise, $\bf L$ and ${\bf R}$ denote the left and 
right multiplication by $\r$. The Bures metric now takes the form
\begin{equation}\label{metric2}
g=\frac{1}{2}\Tr\dr\frac{1}{\bf L+R}\dr\,.
\end{equation}
We set 
$$\chi(t):=\det\left(t\bbox{1}-\r\right)=t^n+k_1t^{n-1}+\dots+k_n,
$$
$k_0:=1$ and $k_i:=0$ for $k>n$  or $k<0$. Hence,
$e_i:=(-1)^i \,k_i$ is the  elementary invariant  of degree 
$i$ and $(-1)^n\chi(t)$ is the characteristic polynomial of $\r$. Moreover, 
let $p_i:=\Tr \r^i$ so that the differential $\dP{i}$ applied to a 
tangent vector $X$ yields
$\dP{i} (X)=i\,\Tr X\r^{i-1}$. Finally, we will 
several times make use of the matrix
\begin{equation}
P:={\renewcommand{\arraystretch}{0.7}
\scriptstyle\left[
\begin{array}{llcl}
p_1&p_2&\dots&p_n\\
p_2&p_3&\dots&p_{n+1}\\
{}^{\vdots}&{}^{\vdots}&{}^{\vdots}&{}^{\vdots}\\
p_n&p_{n+1}&\dots&p_{2n-1}
\end{array}
\right]}\,.
\end{equation}
For instance we have  the 

\smallskip
{\noindent\bf Criterion:} {\it
A  Hermitian $n{\times}n$-matrix $\r\geq0$  is a generic point of ${\cal 
D}$ iff $\det P\neq 0$.}

\smallskip\noindent  
Indeed, $P=V^T\Lambda V$. Therefore, 
$\det P=\det(\Lambda) \det(V)^2=\det(\r)\,\prod_{i<j} 
(\lambda_i-\lambda_j)^2$.\hfill$\Box$

\section{General Formulae}
In order to calculate $g_\r(Y',Y)$ one has to solve the matrix equation
\begin{equation}\label{gl}
\r X+X\r=Y\,.
\end{equation}
Then 2$g_\r(Y',Y)=\Tr Y'X$. The 
matrix (resp.~operator) equation $EX-XF=Y$ was intensively studied
(for a review see \cite{BR97}). A 
basic item is that it has a unique solution $X$ if $E$ and $F$ have 
disjoint spectra (Sylvester, Rosenblum). In our matrix case, $E=\r=-F$,
this is fulfilled because $\r$ is positive. The uniqueness  is also clear, 
not appealing  to this theorem, since we may suppose 
w.l.o.g.~$\r$  to be diagonal with eigenvalues $\lambda_i$ and  equation 
(\ref{gl}) becomes 
$(\lambda_i+\lambda_j)X_{ij}=Y_{ij}$; $i,j=1,\dots n$. 
A further, simple but nice,  tool  in this theory is the use 
of similarities of block matrices, in our case  e.~g.
\begin{equation}\label{sim}
{\renewcommand{\arraystretch}{0.9}
\left[
 \begin{array}{rc}
\bbox{1}&-X\\0 &\bbox{1}
\end{array}
\right]
\left[
\begin{array}{rc}
-\r&Y\\0 &\r
\end{array}
\right]
\left[
\begin{array}{rc}
\bbox{1}&X\\0 &\bbox{1}
\end{array}
\right]=
\left[
\begin{array}{rc}
-\r&0\\0 &\r
\end{array}
\right]\,.}
\end{equation}
Applying  the 
polynomial $\chi$ as an operator function to both sides 
leads in the upper right box to the identity
\begin{equation}\label{lsga}
\chi(-\r)X+M=0\,,
\end{equation}
where $M$ is the upper right box of $\chi$ applied to the above middle 
matrix containing $Y$;
\begin{equation}\label{M1}
M=\sum_{i=1}^n k_{n-i}\sum_{j=0}^{i-1}(-\r)^j Y\r^{i-j-1}\,.
\end{equation}
But  $\chi(-\r)$ is invertible. This can be seen 
as follows. 
The characteristic polynomial  of $-\r$ equals $(-1)^n\chi(-t)$ and the 
positivity of $\r$ implies that $\chi(t)$ and $\chi(-t)$  have no 
common divisors. Hence, by the Euclidean algorithm, there exists two 
polynomials $p,q$ such that
$p(t)\chi(t)+q(t)\chi(-t)=1$, and inserting $\r$ gives 
$q(\r)\chi(-\r)=\bbox{1}$.
Therefore, the solution $X$ is given by 
{\mathletters\label{X}
\begin{equation}\label{X1}
X=-\chi(-\r)^{-1}\sum_{i=1}^n k_{n-i}\sum_{j=0}^{i-1}(-\r)^j Y\r^{i-j-1}\,,
\end{equation}
or, in a more compact form,
\begin{equation}\label{lsg}
{\renewcommand{\arraystretch}{0.9}
{\scriptstyle\left[
 \begin{array}{rc}
-\bbox{1}&X\\0 &0
\end{array}
\right]=-
\left[
\begin{array}{cc}
\chi(-\r)^{-1}&0\\0 &0
\end{array}
\right]\;}
\chi{\scriptstyle\left(\left[
\begin{array}{rc}
-\r&Y\\0 &\r
\end{array}
\right]\right)\,}\,.}
\end{equation}}
The first explicit formula we get for the Bures metric  is:

\smallskip
{\noindent\bf Proposition 1:}
\begin{equation}
g(Y',Y):=-\frac{1}{2}\;\Tr
{\renewcommand{\arraystretch}{0.9}
{\scriptstyle\left[
 \begin{array}{cc}
0&0\\Y' &0
\end{array}
\right]
\left[
\begin{array}{cc}
\chi(-\r)^{-1}&0\\0 &0
\end{array}
\right]\;
}
\chi
{\scriptstyle\left(\left[
\begin{array}{rc}
-\r&Y\\0 &\r
\end{array}
\right]\right)}}\,.
\end{equation}
\hfill$\Box$

\medskip
\noindent
The inverse of $\chi(-\r)$ is again a polynomial expression in
$\r$. Therefore, rewriting  (\ref{X1})   using 
the Cayley-Hamilton theorem,  $X$ can be transformed to
 the form 
$\sum_{1\leq i,j\leq n} a_{ij}\r^{i-1} Y\r^{j-1}$ with exponents less than $n$ 
and coefficients being invariants of $\r$. That means, 
{\mathletters\label{inv}
\begin{eqnarray}\label{inva}
\frac{1}{\bf L+R}&=& 
\sum_{1\leq i,j\leq n} a_{ij}{\bf L}^{i-1} {\bf R}^{j-1}\\
\label{invb}&=&
\left({\bf Id},{\bf L},\dots,{\bf L}^{n-1}\right)A
\left({\bf Id},{\bf R},\dots,{\bf R}^{n-1}\right)^{\rm T}\,,\quad
A:=(a_{i,j})\,.
\end{eqnarray}}
(Of course, this can be seen also directly. Since 
$\bf L+R$ acts on a finite dimensional space, its inverse is a polynomial 
in $\bf L+R$, which again can be reduced to the above form using 
$\chi({\bf L})={\bf L}_{\chi(\r)}=0$ and similar for $\bf R$.)
The solution of (\ref{gl}) given by Smith in \cite{Smith} 
is, essentially, of the above form 
$\sum_{1\leq{i,j}\leq n}a_{ij}\r^{i-1}Y\r^{j-1}$. The 
formulae for the 
coefficients in terms of invariants of $\r$ one  reads off from 
\cite{Smith} will be given at the end of this section. 

The 
representation (\ref{inv}) is unique provided $\r$ is generic. Indeed, if 
there were coefficients such that 
$\sum_{1\leq i,j\leq n} a_{ij}{\bf L}^{i-1}{\bf R}^{j-1}=0$ 
then applying this 
operator to all vectors of a common eigenbasis of $\bf L$ and ${\bf R}$ 
would result in $V^{\rm T}AV=0$. But the Vandermonde matrix $V$ is 
nondegenerate for a generic $\r$ and we would conclude $a_{ij}=0$. 
Moreover, in the generic case the matrix $A$ is necessarily symmetric. 

To get a compact expression for the coefficients in (\ref{inv}) we define
 a $n{\times} n$-matrix $K$ by 
\begin{equation}
\label{K1}K:=
{\renewcommand{\arraystretch}{0.5}
\scriptstyle\left[
\begin{array}{ccccc}
0 & 1 & 0 & \dots & 0\\
0 & 0 & 1 & \dots & 0\\
{}^{\vdots} & {}^{\vdots}&{}^{\vdots}&&{}^{\vdots}\\
0 & 0 & 0 & \dots   & 1\\
-k_n&-k_{n-1}&-k_{n-2}&\dots& -k_1
\end{array}
\right]}\,.
\end{equation}
$K$ carries out 
the reduction of powers of $\r$ by $\chi(\r)=0$. Indeed we have
$$
{\renewcommand{\arraystretch}{0.5}
\scriptstyle\left[
\begin{array}{c}
\r\\ \r^2\\{}^{ \vdots}\\ \r^n
\end{array}
\right]}=
K
{\renewcommand{\arraystretch}{0.5}
\scriptstyle\left[
\begin{array}{c}
\bbox{1}\\ \r\\ {}^{\vdots}\\ \r^{n-1}
\end{array}
\right]}
$$
and similar for the reduction of powers of $\bf L$ and ${\bf R}$.
Thus, multiplying (\ref{invb}) by $\bf L+R$ leads to 
\begin{equation}\label{KA}
K^{\rm T} A +AK=C\,,
\end{equation}
where
\begin{equation}\label{C}
C:={\renewcommand{\arraystretch}{0.5}
\scriptstyle\left[
\begin{array}{cccc}
1&0&\dots&0\\
0&0&\dots&0\\
{}^{\vdots}&{}^{\vdots}&&{}^{\vdots}\\
0&0&\dots&0
\end{array}
\right]}\,
\end{equation}
represents the identity operator ${\bf Id}= {\bf L}^0{\bf R}^0$. 
Note that  $K$ has
the same characteristic polynomial as $\r$.
Now we may proceed as above to find $A$. We replace in (\ref{sim}) 
$-\r$, $\r$, $X$ and $Y$ by $-K^{\rm T}$, $K$, $A$ and $C$ and apply
$\chi(t)$. 
Instead of (\ref{X}) and (\ref{M1}) we obtain
\begin{equation}\label{lsgc}
A=-\chi(-K^{\rm T})^{-1}N\,,
\end{equation}
where 
\begin{eqnarray}\label{Na}
N&=&\nonumber
\chi{\renewcommand{\arraystretch}{0.5}{\scriptstyle\left(\left[
\begin{array}{cc}
-K^{\rm T}&C\\0 &K
\end{array}
\right]\right)}}_{1\,2}
=
\sum_{i=1}^n k_{n-i}\sum_{j=0}^{i-1}(-K^{\rm T})^j CK^{i-j-1}\\
&&\nonumber\\
\label{N}&=&
{\scriptstyle\left[
{\renewcommand{\arraystretch}{0.5}
\begin{array}{ccccc}
k_{n-1}&k_{n-2}&\dots&k_1&1\\
-k_{n-2}&-k_{n-3}&\dots&-1&0\\
{}^{\vdots}&{}^{\vdots}&&{}^{\vdots}&{}^{\vdots}\\
(-1)^{n-1}&0&\dots&0&0
\end{array}}\right]}=
\biggm[(-1)^{i+1}k_{n+1-i-j}\biggm]_{i,j=1}^n\;.
\end{eqnarray}
To see the last equation note that $K^{{\rm T}\,k}CK^j$ has only a $1$ in the 
$(k+1,j+1)$ - position; $K$ moves the $1$ 
coming from $C$ to the right and 
$K^{\rm T}$ moves it down.  
Hence we get 

\smallskip
{\noindent\bf Proposition 2:} {\it The Bures metric equals
\begin{equation}\label{Bures}
g=\frac{1}{2}\sum_{i,j=1}^n a_{ij}\Tr\dr\r^{i-1}\dr\r^{j-1}\,,
\end{equation}
where  $\left(a_{ij}\right)=-\chi(-K^{\rm T})^{-1}N\,$, $K$ and $N$  given by
(\ref{N}) and (\ref{K1}).}\hfill$\Box$

\smallskip
\noindent The probably "most explicit" form of the coefficients $a_{ij}$ 
is given by 

\smallskip
{\noindent\bf Proposition 3} (R.~A.~Smith): {\it 
\begin{equation}\label{Smith}
a_{ij}=
\frac{(-1)^i}{2\, \det H}
\sum_{r=0}^{n-i}\sum_{s=0}^{n-j}(-1)^r k_r k_s \Phi(\frac{i+j+r+s}{2})\,,
\end{equation}
where 
\begin{equation}\label{H}
H=
{\scriptstyle\left[
{\renewcommand{\arraystretch}{0.5}
\begin{array}{cccc}
k_1&k_3&\dots&k_{2n-1}\\
k_0&k_2&\dots&k_{2n-2}\\
{}^{\vdots}&{}^{\vdots}&{}^{\vdots}&{}^{\vdots}\\
0&0& \dots&k_n
\end{array}}\right]}=
\bigm[k_{2j-i}\bigm]_{i,j=1}^n
\end{equation}
and $\Phi(m)=0$ if $m$ is not an integer, and otherwise $\Phi(m)$
is the cofactor in $\det H$ of the element in the first row and $m$-th 
column of $H$.\hfill$\Box$
}

\smallskip
{\noindent\bf Remark}: {\it The determinant of $H$ is not equal 
zero, more precisely:}
$$
\det H=
(-1)^\frac{n(n+1)}{2} 
\prod_i \lambda_i\;\prod_{i<j} (\lambda_i+\lambda_j)
\neq 0. 
$$

\smallskip
\noindent Indeed, changing the order of rows and columns we get
$\det H= (-1)^\frac{n(n+1)}{2}\det\left[e_{n+1-2i+j}\right]$.
But the last determinant is just the (symmetric) Schur function 
of the eigenvalues of $\r$ 
(comp.~\cite{Donald}, I.3) related to
the partition $(n,n-1,\dots,1)=(1,\dots,1)+(n-1,\dots,0)$ 
leading to the above product.
\hspace{1cm}\hfill$\Box$
\section{A Formula Adapted to 
${\cal D}\approx{\Bbb R}^{\lowercase{n}}\times
 {}^{{\rm U}({\lowercase{n}})}/_{{\rm T}^{\lowercase{n}}}$}
Every $\r\in{\cal D}$ can be diagonalized with a suitable 
unitary $u$;
$$\r=u\mu^2 u^*\;,\qquad\mu=
{\rm diag}(\mu_1,\dots,\mu_n)\;,\quad\mu_i\in{\Bbb R}_+$$
and we have 
$\dr=2u\,\mu\dmu u^*+u\,[u^*\du,\mu^2]\,u^*$. 
The inverse problem now splits into
$$
\frac{1}{{\bf L}_\r+{\bf R}_\r}\,(\dr)=
u\,\left(\,\mu^{-1}\dmu+
\frac{1}{{\bf L}_{\mu^2}+{\bf 
R}_{\mu^2}}\,([\,u^*\du,\mu^2\,])\,\right)\,u^*\,
$$
and we get for the metric
\begin{equation}
g_\r=
\Tr \dmu\dmu+
\frac{1}{2}\Tr[\,u^*\du,\mu^2\,]
\frac{1}{{\bf L}_\mu^2+{\bf R}_\mu^2}\,([\,u^*\du,\mu^2\,])\,.
\end{equation}
Therefore, in a neighbourhood of a generic point  the Riemannian 
manifold $\cal D$ locally looks like 
${\Bbb R}^n\times{}^{{\rm U}(n)}/_{{\rm T}^n}$,
where ${\Bbb R}^n$ is equipped with the standard metric and 
the metric on the homogeneous space 
${}^{{\rm U}(n)}/_{{\rm T}^n}$
depends on the first parameter. 
The tangent space at $\r$ splits into
\begin{equation}\label{deco}
{\rm T}_\r{\cal D}={\rm T}_\r^\shortparallel+
{\rm T}_\r^{\shortparallel\,{\scriptstyle\perp}}\,,
\end{equation}
where 
${\rm T}_\r^{\scriptstyle\shortparallel}$
is the subspace of Hermitian 
matrices commuting with $\r$. Its orthogonal complement (w.~r.~to the Bures 
metric) is the space of all $[a,\r]$, $a$ - antihermitian. If $\r$ is 
diagonal then (\ref{deco}) is  the decomposition  into  
diagonal and off-diagonal  Hermitian matrices.

From now on let $\r$ be a generic point. By ${\bf P}$
and ${\bf P}^{\scriptstyle\perp}={\bf Id-P}$ we denote  the (orthogonal) 
projectors onto the subspaces in (\ref{deco}). 

\smallskip
{\noindent\bf Lemma:}
\begin{mathletters}\label{P}
\begin{eqnarray}
{\bf P}(Y)&=&\label{P1}
\sum_{i,j=1}^n\r^i
\,\left(P^{-1}\right)_{ij}\,
\Tr Y\r^{j-1}\\
&=&\label{P2}
\sum_{i,j=1}^n
\,\left(P^{-1}\right)_{ij}\,
\r^i Y\r^{j-1}\,.
\end{eqnarray}
\end{mathletters}

\smallskip
{\noindent\bf Proof:} To show  (\ref{P1}) we use that for a generic 
$\r$ 
the powers $\r,\r^2,\dots,\r^n$ form a basis of the vector space of all 
Hermitian matrices commuting with $\r$. Moreover, 
$4\,P_{ij}=g_\r(\r^i,\r^j)$ and $4\,\Tr Y\r^{j-1}=g_\r(Y,\r^j)$. 
Having this in mind (\ref{P1}) is just the usual  formula for the 
orthogonal projection onto a subspace with a given basis.
To see (\ref{P2}) let $Y_{\alpha\beta}$, $\alpha,\beta=1,\dots,n$
be a common eigenbasis of $\bf L$ and ${\bf R}$,  
$\r Y_{\alpha\beta}=\lambda_\alpha Y_{\alpha\beta}$, 
$Y_{\alpha\beta}\,\r=\lambda_\beta Y_{\alpha\beta}$ 
($Y_{\alpha\beta}$ - may not be Hermitian). 
Then the  complex span of ${\rm T}_\r^{\scriptstyle\shortparallel}$
resp.~${\rm T}_\r^{\scriptstyle\shortparallel\perp}$
is generated by all $Y_{\alpha\beta}$ with
$\alpha=\beta$ resp.~$\alpha\neq\beta$. 
For 
$Y=Y_{\alpha\beta}$ the right hand 
side of (\ref{P2}) yields 
$\eta_{\alpha\beta}Y_{\alpha\beta}$, where 
$$\eta_{\alpha\beta}=
\sum_{i,j=1}^n\left(P^{-1}\right)_{ij}\,\lambda_\alpha^i\lambda_\beta^{j-1}=
\bigg(\Lambda\,
V\,P^{-1}\,V^T\bigg)_{\alpha\beta}\,,
$$ 
$V$ - the Vandermonde matrix of eigenvalues of $\r$. But
$P=V^T\,\Lambda\,V$
implies $\eta_{\alpha\beta}=\delta_{\alpha\beta}$. 
\hfill$\Box$

\smallskip
\noindent 
From (\ref{P})  we now get
\begin{eqnarray*}
\frac{1}{\bf L+R}\,{\bf P}(Y)&=&
\frac{1}{2}\sum_{i,j=1}^n\r^{i-1}
\,\left(P^{-1}\right)_{ij}\,
\frac{\dP{j}}{j}(Y)\,,\\
\frac{1}{\bf L+R}\,{\bf P}^{\scriptstyle\perp}(Y)&=&
\sum_{i,j=1}^n
\,\left\{a_{ij}-\frac{1}{2}\left(P^{-1}\right)_{ij}\right\}\,
\r^{i-1} Y\r^{j-1}\,,
\end{eqnarray*}
where we used 
$\Tr j Y\r^{j-1}=\dP{j}(Y)$ and 
$\frac{1}{\bf L+R}(Y)=\frac{1}{2}\r^{-1}Y$ for 
$Y\in {\rm T}^\shortparallel$.
The matrix $\left(a_{ij}\right)$ 
is given by Proposition 2 or 3.
Inserting these equations into
$$g=\frac{1}{2}\Tr\left(
{\bf P}(\dr)\frac{1}{\bf L+R}\,{\bf P}(\dr)+
\dr \frac{1}{\bf L+R}\,{\bf P}^{\scriptstyle\perp}(\dr)
\right) 
$$
yields

\smallskip
{\noindent \bf Proposition 4:} {\it
The decomposition  
$\displaystyle g=g_{\restriction_{\rm T^\shortparallel}}+
g_{\restriction_{\rm T^{\shortparallel\scriptstyle\perp}}}$ 
of the Bures metric is given by
\begin{equation}\label{BURES}
g=\frac{1}{4}\sum_{i,j=1}^n\frac{\dP{i}}{i}
  \left(P^{-1}\right)_{ij}
  \frac{\dP{j}}{j}\;+\;
  \frac{1}{4}\sum_{i,j=1}^n \left(2a_{ij}-\left(P^{-1}\right)_{ij}\right)
\Tr\dr\r^{i-1}\dr\r^{j-1}\,.
\end{equation}
\hfill$\Box$ }
\section{Examples}
Propositions 1-4 involve elementary and power invariants, which can be 
expressed by each other (comp.\cite{Donald}). Rewriting the identity  
$$
m k_m+\sum_{r=1}^{m}p_rk_{m-r}=0\,,\quad m=1,2,\dots
$$
as  a system of linear equations for the $e_i$-s resp.~the  $p_i$-s
one gets the  relations
$$
e_i=\frac{1}{i!}\det
{\renewcommand{\arraystretch}{0.7}
\scriptstyle
\left[
\begin{array}{ccccc}
p_1&1&0&\dots&0\\
p_2&p_1&2&\dots&0\\
\vdots&\vdots&\vdots&&\vdots\\
p_{i{-}1}&p_{i{-}2}&p_{i{-}3}&\dots&i{{-}}1\\
p_i&p_{i{-}1}&p_{i{-}2}&\dots&p_1
\end{array}
\right]}\,,\qquad
p_i=\det
{\renewcommand{\arraystretch}{0.7}
\scriptstyle
\left[
\begin{array}{ccccc}
e_1&1&0&\dots&0\\
2e_2&e_1&1&\dots&0\\
\vdots&\vdots&\vdots&&\vdots\\
(i{-}1)e_{i{-}1}&e_{i{-}2}&e_{i{-}3}&\dots&1\\
ie_{i}&e_{i{-}1}&e_{i{-}2}&\dots&e_1
\end{array}
\right]}\,.
$$
This identity also allows for expressing  the power invariants $p_m$, $m>n$,
by invariants of degree less or equal $n$. Especially the $i+1$ - row of 
our matrix $P$ equals $(p_{i+1},\dots,p_{n+i})= (p_{1},\dots,p_{n})K^i$.

The number of terms in Propositions 1-4  rapidly increases 
with the dimension $n$. Thus we give 
only  certain expressions for $n=2,3$ 
in terms of power resp.~elementary invariants.
Concerning Proposition 1 we get
$$
g(Y',Y):=\\
\frac{1}{2}\left\{
\begin{array}{lcl}
\Tr Y'(\r^2+e_1\r+e_2\bbox{1})^{-1}(\r Y-Y\r+e_1Y)&\mbox{for}&n=2\\
\Tr Y'(\r^3+e_1\r^2+e_2\r+e_3\bbox{1})^{-1}\times\\
\qquad(\r^2 Y-\r Y\r+Y\r^2 +e_1(\r Y- Y\r)+e_2Y)&\mbox{for}&n=3
\end{array}\right.
$$
The following terms appear in Propositions 2-4:

\smallskip\noindent$\underline{n=2}$:
\begin{eqnarray*}
g_{\restriction_{\rm T^\shortparallel}}&=&
\frac{1}{4(p_1p_3-p_2^2)}
\left(\begin{array}{ccc}
\frac{\dP{1}}{1}&,&\frac{\dP{2}}{2}
\end{array}\right)
\left[\begin{array}{cc}
p_3&-p_2\\-p_2&p_1
\end{array}\right]
\left(\begin{array}{c}
\frac{\dP{1}}{1}\\\frac{\dP{2}}{2}
\end{array}\right)\\
&=&
\frac{1}{4e_2(e_1^2-4e_2)}
\left(\begin{array}{ccc}
\de{1}&,&\de{2}
\end{array}\right)
\left[\begin{array}{cc}
e_1e_2&-2e_2\\-2e_2&e_1
\end{array}\right]
\left(\begin{array}{c}
\de{1}\\\de{2}
\end{array}\right)\\
\end{eqnarray*}
$$
A=\frac{1}{2e_1e_2}
\left[
\begin{array}{cc}
e_1^2+e_2&-e_1\\-e_1&1
\end{array}\right]
$$
$$
2A-P^{-1}=
\frac{2}{e_1(e_1^2-4e_2)}
\left[
\begin{array}{lr}
-2e_2 & e_1 \\
e_1 & -2 
\end{array}
\right]\\
=
\frac{2}{p_1(2p_2-p_1^2)}
\left[
\begin{array}{lr}
p_2-p_1^2&p_1\\
p_1 & -2 
\end{array}
\right]
$$

\smallskip\noindent$\underline{n=3}$:
$$
\det P=-p_3^3 + 2p_2 p_3 p_4 - p_1 p_4^2 - 
  p_2^2 p_5 + p_1 p_3p_5\\
$$
$$
P^{-1}=\frac{1}{\det P}
\left[
\begin{array}{ccccc}
p_3p_5-p_4^2&&
p_3p_4-p_2p_5&&
p_2p_4-p_3^2\\
\ast &&
p_1p_5-p_3^2&&
p_2p_3-p_1p_4\\
\ast &&\ast &&
p_1p_3-p_2^2
\end{array}
\right]
$$
$$
A=\frac{1}{2e_3(e_1e_2-e_3)}
\left[
\begin{array}{ccccc}
e_1e_2^2+e_1^2e_3-e_2e_3&&
-e_1^2e_2&&e_1e_2-e_3\\
\ast&&e_1^3+e_3&&-e_1^2\\
\ast&&\ast&&e_1
\end{array}
\right]
$$
\acknowledgments
I would like to thank  A.~Uhlmann 
for stimulating discussions and P.~Alberti and  K.~Schm\"udgen for a
valuable advice.


\begin{thebibliography}{*}
\bibitem{Uh92b}
A.~Uhlmann: The metric of Bures and the geometrical phase,
in: {\it Quantum Groups and Related Topics}, 
 eds.~R.~Gielerak et al., Kluwer Acad.~Publishers, 1992, pp. 267--274

\bibitem{UD}J.~Dittmann and A.~Uhlmann:  Connections and metrics respecting 
standard purification, quant-ph/9806028

\bibitem{Hu92b}
M.~H\"ubner: Explicit computation of the Bures distance for density matrices,
Phys.~Lett. {\bf A 163} (1992), 239--242

\bibitem{Di93a}
J.~Dittmann: On the Riemannian geometry of finite dimensional mixed states,
Sem.~S.~Lie {\bf 3} (1993), 73--87

\bibitem{Twam}J.~Twamley: Bures and statistical distance for 
sqeezed thermal states, J.~Phys.~{\bf A 29} (1996), 3723 

\bibitem{Slater}P.~B.~Slater: Bures Metrics for Certain High-Dimensional 
Quantum Systems, Phys.Lett. {\bf A 244 }(1998), 35-42

\bibitem{BR97}
R.~Bhatia and P.~Rosenthal: How and  why to solve the operator 
equation $AX-XB=Y$, Bull.~London Math.~Soc. {\bf 29} (1997), 1--21


\bibitem{Smith}R.~A.~Smith: Matrix Calculations for Liapunov Quadratic 
Forms, J.~Differential Equations {\bf 2} (1966), 208--217

\bibitem{Donald} I.~G.~Macdonald,
{\it Symmetric Functions and Hall 
Polynomials}, Oxford University Press Inc., New York, 1995


\end{thebibliography}
\end{document}